\documentclass[11pt]{article}

\usepackage{authblk}

\usepackage{graphicx}

\usepackage{float}

\makeatletter
\newcommand{\printfnsymbol}[1]{%
  \textsuperscript{\@fnsymbol{#1}}%
}
\makeatother

\begin{document}
\title{Bayesian inference of set-point viral load transmission models}
\author[1,2]{Pieter Libin}
\author[1]{Laurens Hernalsteen}
\author[2]{Kristof Theys}
\author[3,4]{Perpetua Gomes}
\author[5]{Ana Abecasis}
\author[1]{Ann Now\'{e}}
\affil[1]{Artificial Intelligence lab, Department of computer science, Vrije Universiteit Brussel}
\affil[2]{Department of Microbiology and Immunology, Rega Institute for Medical Research, KU Leuven - University of Leuven, Leuven, Belgium}
\affil[3]{Laboratório de Biologia Molecular, LMCBM, SPC.HEM, Centro Hospitalar Lisboa Ocidental, Lisboa, Portugal}
\affil[4]{Centro de Investigação Interdisciplinar Egas Moniz, CiiEM, ISCSEM, Almada, Portugal}
\affil[5]{Global Health and Tropical Medicine, GHTM, Instituto de Higiene e Medicina Tropical, IHMT, Universidade Nova de Lisboa, UNL, Lisboa, Portugal}
\maketitle              % typeset the header of the contribution
\begin{abstract}
When modelling HIV epidemics, it is important to incorporate set-point viral load and its heritability. As set-point viral load distributions can differ significantly amongst epidemics, it is imperative to account for the observed local variation. This can be done by using a heritability model and fitting it to a local set-point viral load distribution. However, as the fitting procedure needs to take into account the actual transmission dynamics (i.e., social network, sexual behaviour), a complex model is required. Furthermore, in order to use the estimates in subsequent modelling analyses to inform prevention policies, it is important to assess parameter robustness.

  In order to fit set-point viral load models without the need to capture explicitly the transmission dynamics, we present a new protocol. Firstly, we approximate the transmission network from a phylogeny that was inferred from sequences collected in the local epidemic. Secondly, as this transmission network only comprises a single instance of the transmission network space, and our aim is to assess parameter robustness, we infer the transmission network distribution. Thirdly, we fit the parameters of the selected set-point viral load model on multiple samples from the transmission network distribution using approximate Bayesian inference. 

 Our new protocol enables researchers to fit set-point viral load models in their local context, and diagnose the model parameter's uncertainty. Such parameter estimates are essential to enable subsequent modelling analyses, and thus crucial to improve prevention policies.

\end{abstract}
\section{Introduction}
\begin{figure}[H]
  \centering
  \includegraphics[width=\textwidth]{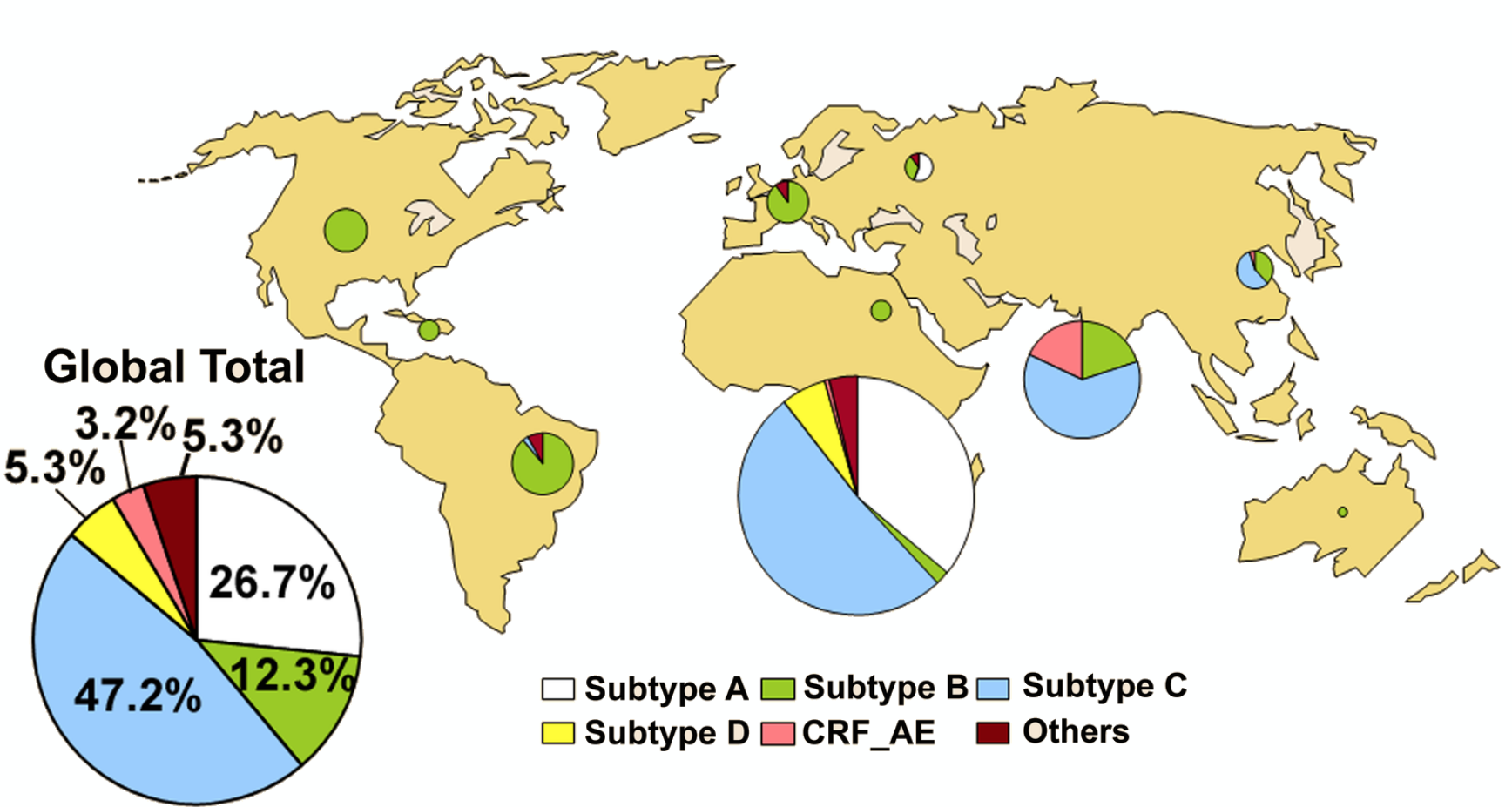}
  \caption{Global map showing the spread of HIV subtypes per continent (Wikipedia, derived from 
	\protect\cite{osmanov2002estimated}).}
  \label{fig:hiv_types_map}
\end{figure}

The human immunodeficiency virus (HIV) targets the immune system of the host and eventually leads to acquired immunodeficiency syndrome. There are two different types of HIV: HIV-1 and HIV-2. HIV-1 is divided into four groups: M, N, O, P. HIV-1 group M is responsible for the current global HIV epidemic, and has been subdivided  into subtypes A, B, C, D, F, G, H, J and K \cite{Lemey2006}. Different geographical regions are associated with particular subtype distributions \cite{osmanov2002estimated}, as visualized in Figure \ref{fig:hiv_types_map}.

In 2017, 36.9 million people worldwide were living with HIV-1 and 1.8 million new HIV infections occurred \cite{who2017hivaids}. Although this number of new infections per year is decreasing thanks to easier access to treatment, this decrease is stagnating and HIV remains a great concern, due to scale-up of antiretroviral treatment, especially in sub-Saharan Africa \cite{wang2016estimates}.

In 2016, a total of 29444 new HIV diagnoses were reported by the 31 countries of the EU/EEA \cite{cdc_hiv_report_2017}.
In Portugal, over 44 624 people are currently living with HIV, according to the national report of 2017 \cite{national_port_2017}.

HIV is no longer an untreatable condition and an HIV therapy consists of administering a drug regimen, an highly active anti-retroviral therapy. HIV is a rapidly replicating virus, because of this, HIV can become resistant to the administered therapy. To mitigate the effects of HIV-1 drug resistance, it is important to closely follow up treated patients.

In order to closely monitor the diseases progression of HIV-1 patients, the viral load (i.e., the number of viral particles in the serum) needs to be measured regularly. Closely related to this marker is the set-point viral load, i.e., the viral load that is measured after the acute phase of the infection (see Figure \ref{fig:untreated_viralload_cd4_function}).
An individual's set-point viral load is associated with disease progression and transmissibility of the virus, as such, the distribution of the set-point viral loads in a population is an important determinant of the transmission dynamics of the epidemic. Furthermore, as set-point viral load is partly inherited by the infectee, this induces a complex interplay between intra-patient and inter-patient processes. Additionally, it has been well studied that set-point viral load distributions can differ significantly between different locations \cite{fraser2007variation}. It is therefore important to consider the population's set-point viral load distribution to better understand the characteristics of a particular HIV-1 epidemic.

\begin{samepage}
\begin{figure}[H]
  \centering
  \includegraphics[width=0.8\textwidth]{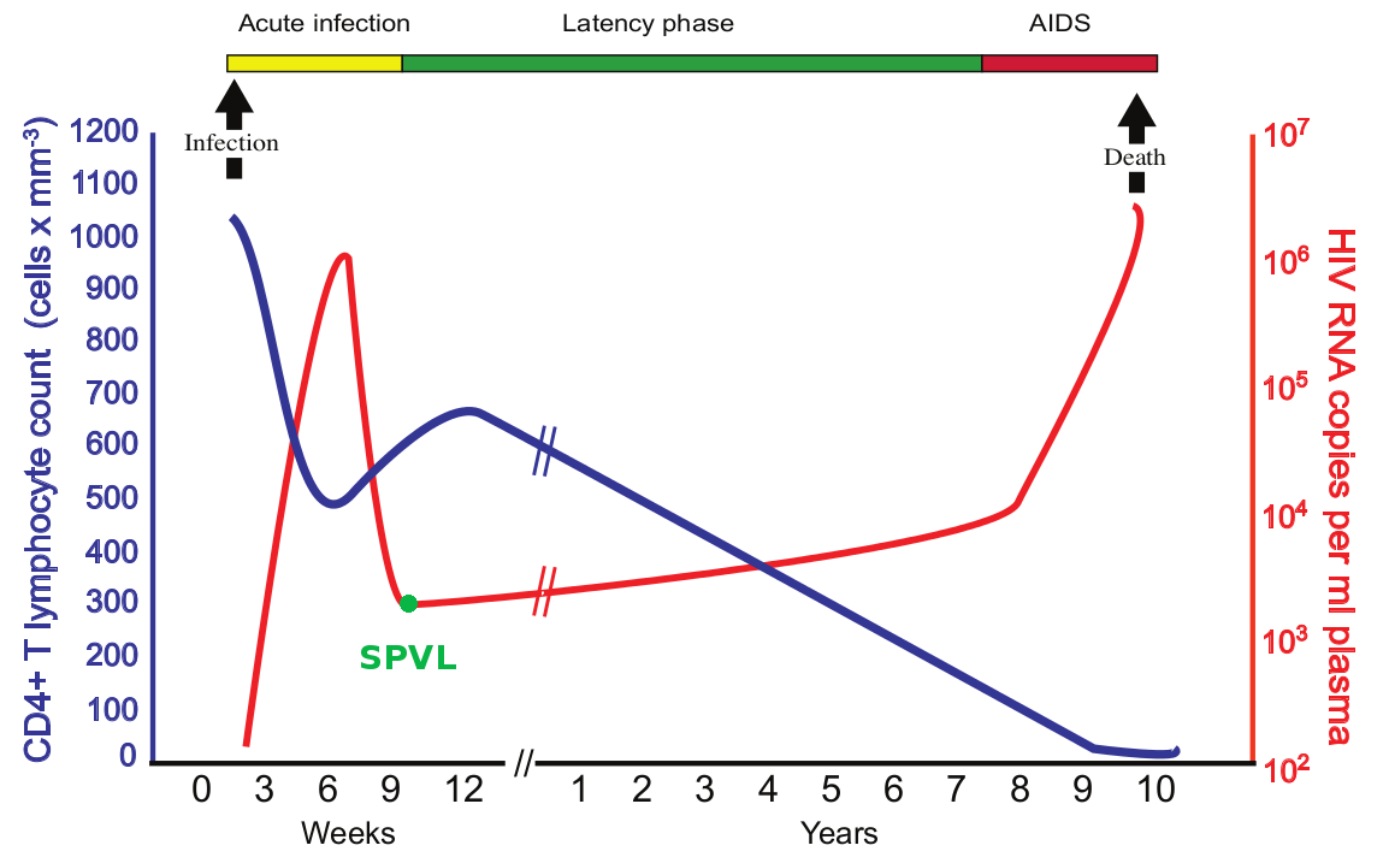}
  \caption[Plot of the average viral load function (red) and average CD4 cell count (blue) of an untreated HIV patient.] 
	{Plot of the average viral load function (red) and average CD4 cell count (blue) of an untreated HIV patient. 
	The set-point viral load is the viral load that directly follows the acute phase, at around 12 weeks after seroconversion. This figure was adapted from the figure presented in \protect\cite{theys2010hiv}.}
  \label{fig:untreated_viralload_cd4_function}
\end{figure}
\end{samepage}

As such, when modelling HIV epidemics, it is important to incorporate set-point viral load and its heritability \cite{theys2018impact}. As set-point viral load distributions can differ significantly amongst epidemics, it is imperative to account for the observed local variation. This can be done by using a heritability model and fitting it to a local set-point viral load distribution. However, as the fitting procedure needs to take into account the actual transmission dynamics (i.e., social network, sexual behaviour), a complex model is required. Furthermore, in order to use the estimates in subsequent modelling analyses to inform prevention policies, it is important to assess the parameter robustness.

\section{Objectives}
Our aim in this work is to provide a new protocol, to fit set-point viral load models without the need to explicitly capture the transmission dynamics. We will present the methodology for this new protocol and evaluate it in the context of the Portuguese HIV-1 epidemic. The Portuguese HIV-1 epidemic is unique, as it comprises two parallel epidemics, clearly distinguished by their HIV-1 subtype, i.e., HIV-1 subtype B and HIV-1 subtype G \cite{Palma2007}.

Our new protocol will enable researchers to fit set-point viral load models in their local context (i.e., the specifics of the epidemic: virus' evolutionary rate, diagnosis rate, transmission route, ...), and diagnose the model parameter's uncertainty (see Figure \ref{fig:posteriors} and Figure \ref{fig:posteriors_ln}). Such parameter estimates are essential to enable subsequent modelling analyses and therefore crucial to improve prevention policies.

\section{Methods}
\begin{figure}[h]
  \centering
  \includegraphics[width=0.4\linewidth]{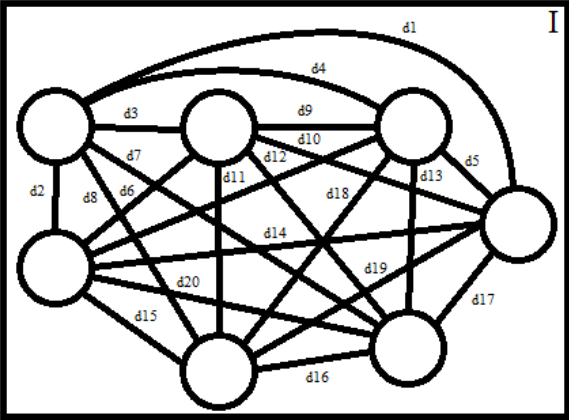}
  \includegraphics[width=0.4\linewidth]{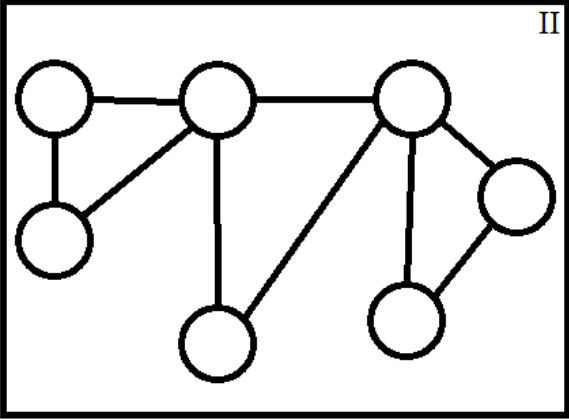}
  \includegraphics[width=0.4\linewidth]{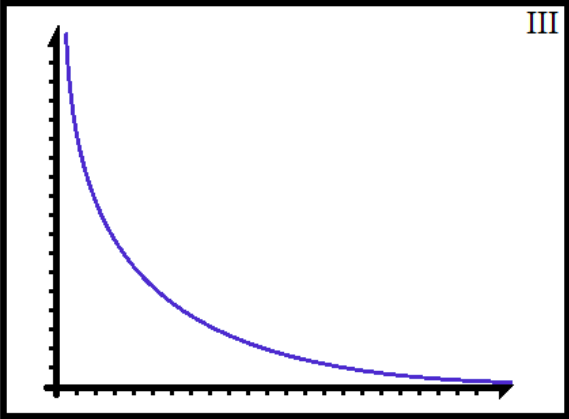}
  \includegraphics[width=0.4\linewidth]{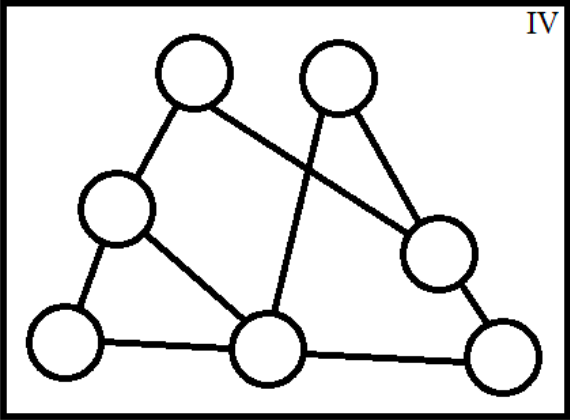}
\caption{A visualization of the work flow of our protocol: 
computing the pairwise genetic distances between viral isolates (I), removing all 
connections whose distance is over a certain threshold (II), fitting a neighbor 
degree distribution on this network (III) and then sampling a network from this 
distribution (IV).
\label{fig:workflow_1}}
\end{figure}

\begin{figure}[h]
\centering
\includegraphics[width=0.4\linewidth]{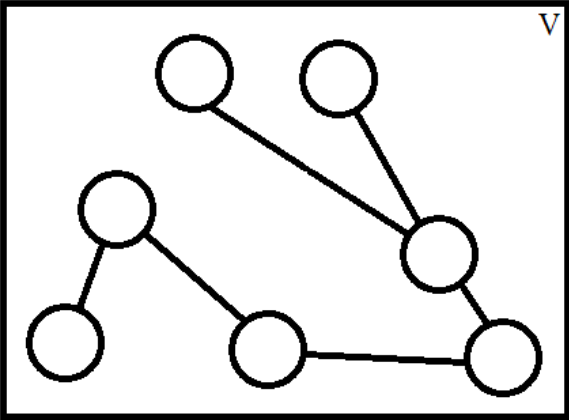}
\includegraphics[width=0.4\linewidth]{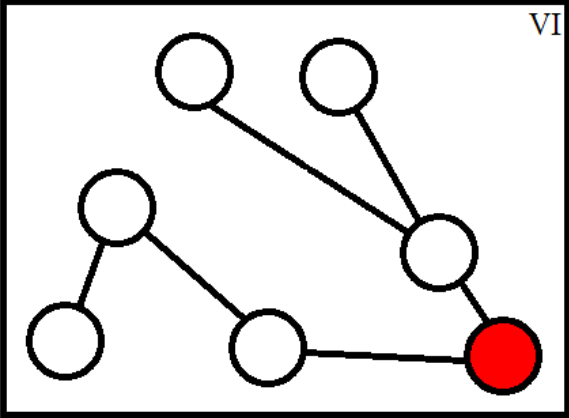}
\includegraphics[width=0.4\linewidth]{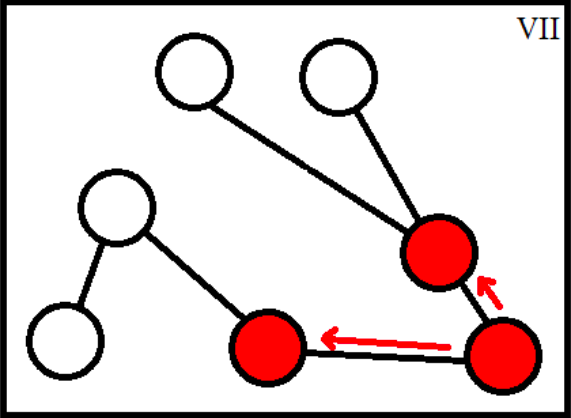}
\includegraphics[width=0.4\linewidth]{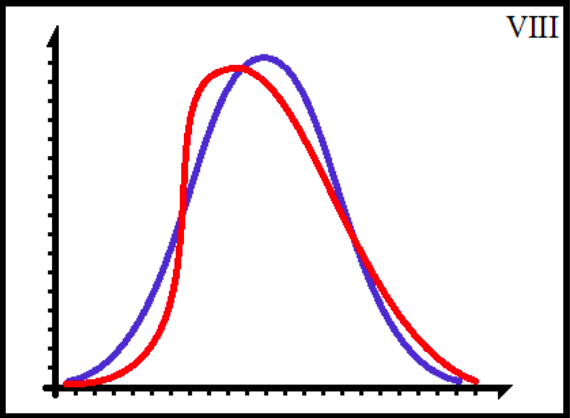}
\caption{A visualization of the protocol work flow: 
sampling a spanning tree from the network of potential transmission partners (V), 
seeding a random node with a set-point viral load value (VI), propagating the infection 
through the network (VII) and comparing the generated SPVL distribution to the empirical 
SPVL distribution to obtain a quality measure of the model parameters used for 
seeding and propagating the SPVLs (VIII).
\label{fig:workflow_2}}
\end{figure}

In Figure \ref{fig:workflow_1}, we give an overview of the procedure to infer the transmission network distribution. In Figure \ref{fig:workflow_2}, we illustrate how we infer the set-point viral load (SPVL) transmission model.

First, we calculate the genetic distances between the patient's genetic virus sequences. This distance matrix, can be represented as a graph where nodes represent patients and the edges between the nodes are weighted with the genetic distance between the genetic virus sequences of these patients (I). 
We then remove all edges from this graph where the distance exceeds a specified threshold. This results in a network where every node is connected to their potential transmission partners (II) \cite{wertheim2013global}.
We then infer the degree distribution of the network of potential transmission partners (III).
This model is used to sample possible transmission networks of the transmission network distribution (IV).
We sample the network using Chung's algorithm \cite{chung2002connected}.

%\begin{minipage}{0.47\textwidth}
%\end{minipage} \hfill
%\begin{minipage}{0.5\textwidth}

%\end{minipage}

%\begin{minipage}{0.47\textwidth}
%\end{minipage} \hfill
%\begin{minipage}{0.5\textwidth}

From this sampled network of potential transmission partners, we obtain a minimum spanning tree (V).
We will use this spanning tree to model the HIV transmission network. Every node in the graph represents a patient. In this spanning tree, every edge represents an actual HIV transmission. We then seed random nodes with a set-point viral load (VI, the infected node is shown in red). Consequently, we use a model for transmitting set-point viral load values to infect the neighbors of this node 
(VII). We propagate the infection through the entire network until every node has been infected and thus was assigned a set-point viral load value. We extract the distribution of generated set-point viral loads from the network and compare this to the empirical distribution estimated from the viral loads extracted from the Portuguese HIV resistance database (VIII) \cite{theys2009rise}. We use Approximate Bayesian Computation in combination with the previous steps to learn the parameter distribution of the set-point viral load model.

%\end{minipage}

\subsection{Transmission networks as graphs}
\begin{figure}[H]
\centering
\includegraphics{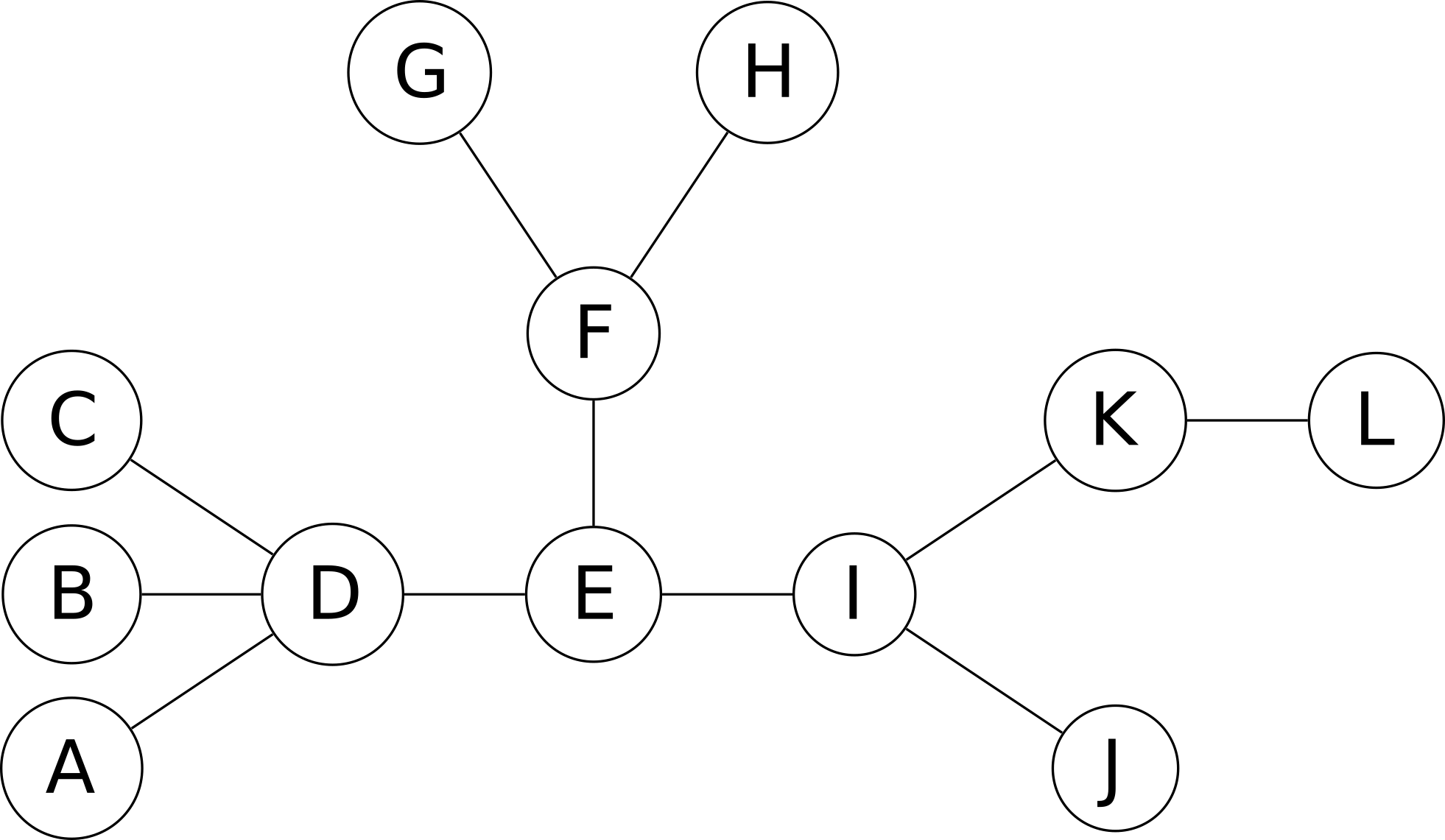}
\caption{An example HIV transmission network represented as an undirected graph. The undirectedness means it is impossible to see whether patient D has infected patient E or vice versa.}
\label{fig:example_transmission_network}
\end{figure}

The topology of a network of HIV-infected individuals is often visualized as a phylogenetic tree.
However, a phylogenetic tree depicts the distinct viral lineages rather than 
the relation between an infector and an infectee \cite{wertheim2013global,pillay2007hiv}.
We therefore represent transmission networks as graphs \cite{wertheim2013global}, where every node represents a patient and every edge between nodes encodes that those patients are possible transmission partners. An example of a graph is given in Figure \ref{fig:example_transmission_network}. 
For this example, the graph represents that patients A, B and C were possibly infected by patient D. 
However, since transmission networks are undirected graphs, there are multiple possible interpretations. It could, for example, be possible that patient A was the one to infect patient D, who then infected patients B and C.

In order to approximate the transmission network, we use the distance matrix, where each entry $(i,j)$ comprises the distance between the genetic virus sequences of patient $p_i$ and $p_j$. From this distance matrix, we create a fully connected graph in which each node represents a patient and the edge between every two nodes is the genetic distance between the patients' virus sequences 
(I in Figure \ref{fig:workflow_1}). We then remove all edges with a weight exceeding a specified threshold (II in Figure \ref{fig:workflow_1}). Every edge now represents a possible transmission. Subsequently, we compute a spanning tree from this network to obtain a realistic HIV transmission network (III in Figure \ref{fig:workflow_1}).

\begin{figure}[H]
\centering
  \includegraphics[width=0.4\linewidth]{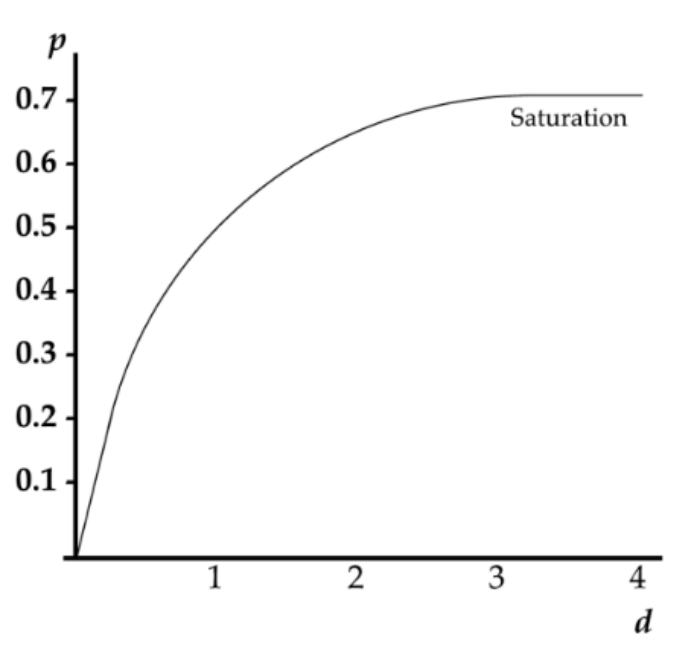}
  
\caption{A plot of the true distance $d$ versus the $p$-distance illustrates the 
	saturation of the distance measure for high divergence \protect\cite{lemey2009phylogenetic}.
\label{fig:saturation_plot}}
\end{figure}

The simplest genetic distance is the pairwise distance measure, that corresponds to the proportion of changed nucleotide sites to the total number of nucleotide sites \cite{lemey2009phylogenetic}. However, the pairwise distance measure underestimates the actual evolutionary distance, as differences at nucleotide sites might get reverted during the evolutionary process, or we might miss evolutionary signal between the observed nucleotide sites (show in Figure \ref{fig:saturation_plot}). In order to mitigate this bias, different nucleotide substitution models were developed. In this work, we use the Tamura-Nei model \cite{tamura1993estimation}, a substitution model that allows for different nucleotide substitution rates, as this has previously been shown adequate to infer transmission networks \cite{wertheim2013global}.

As this transmission network only comprises a single instance of the transmission network space, and our aim is to assess the parameter's robustness, we infer the transmission network distribution. We model this using a Waring degree distribution $\mathit{Waring}(x, \rho)$ or a Yule degree distribution $\mathit{Yule}(x, \rho)$ \cite{wertheim2013global}, for which we need to estimate the $\rho$ parameter based on the transmission network that was inferred from the virus sequence data.

\begin{figure}[H]
\centering
	\includegraphics[width=0.6\linewidth]{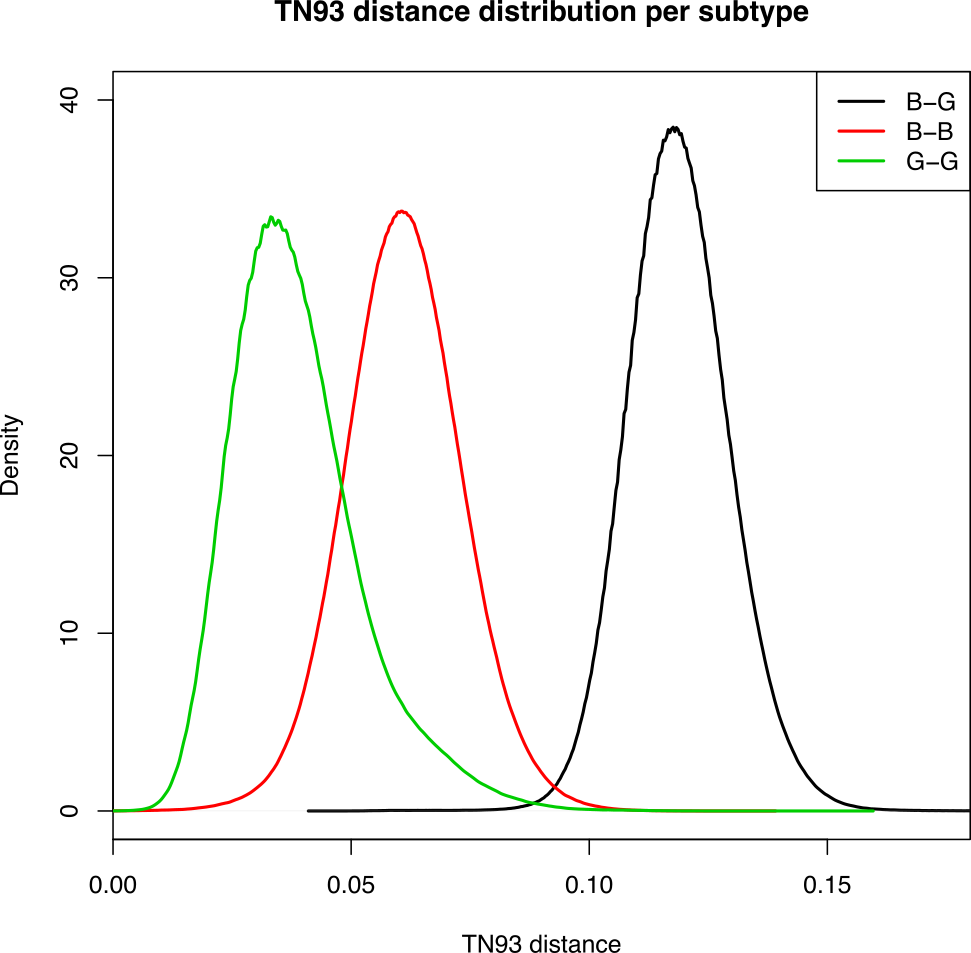}
        \caption{
The distance distribution of our data is shown per combination of subtypes where we show the density of the distance distribution. It is clear that the distance between patients with subtype G and patients with subtype B is significantly larger than the distance between two 
patients with the same subtype.
\label{fig:genetic_distance_per_subtype}}
\end{figure}

In \cite{wertheim2013global}, $\rho$ was estimated on the complete set of sequences. We do however make the observation that the genetic distance distribution of HIV-1 subtype B and HIV-1 subtype G sequences  significantly differs, as is visualized In Figure \ref{fig:genetic_distance_per_subtype}. In our experiments, we will therefore estimate a separate distribution for the different subtypes.

\subsection{Models of HIV spread}
\label{sec:models_of_hiv_spread}

Research has shown that the value of a patient's set-point viral load is partially caused by the genetic code of the virus and is thus partially transmitted 
\cite{hodcroft2014contribution,fraser2007variation,alizon2010phylogenetic,hool2013virus}. 
To build a transmission network for investigating the distribution of set-point viral loads that arise, two methods for establishing a new set-point viral load must be applied. Firstly, a method for assigning an SPVL to a patient zero in a network (we call this seeding):
\begin{equation}
		P(\textbf{spvl})
\end{equation}
Secondly, a method for calculating the SPVL of an infectee, given the SPVL of the infector:
\begin{equation}
  P(\textbf{spvl}_{\mathit{infectee}} | \textbf{spvl}_{\mathit{infector}})
\end{equation}

The Simpact model \footnote{https://github.com/j0r1/simpactcyan} provides two  models to determine new set-point viral loads. The first model uses a two-dimensional probability density function to determine the probability of having a set-point viral load for every viral load of the infectee and the infector. The second model uses a one-dimensional probability density function to seed new viral loads and uses the set-point viral load of the infector with noise for transmitted set-point viral loads.

The first model uses a two-dimensional distribution to calculate the probability of an infected person (infectee) having a certain set-point viral load, given the set-point viral load of the person that infects (infector). The conditional probability function we are interested in is

\begin{equation}
		P(\textbf{spvl} _{\mathit{infectee}} |  \textbf{spvl} _{\mathit{infector}})
\end{equation}

which is implemented in this model as a joint probability function

\begin{equation}
		P(\textbf{spvl} _{\mathit{infector}} ,  \textbf{spvl} _{\mathit{infectee}})
\end{equation}

To seed a new set-point viral load without knowledge of the set-point viral load of the infector, the marginal distribution is used:

\begin{equation}
		P(\textbf{spvl} _{\mathit{infectee}}) = \int P(\textbf{spvl} _{\mathit{infector}} ,  \textbf{spvl} _{\mathit{infectee}}) d\textbf{spvl}_{\mathit{infectee}}
\end{equation}

The actual two-dimensional probability density function from the Simpact study is defined by a bi-variate normal distribution with equal parameters. The standard case of a bi-variate normal distribution is defined by:
\begin{equation}
		P(x, y) = \frac{1}{2\pi\sigma_{1}\sigma_{2}\sqrt{1 - \rho^2}} exp \lbrack- \frac{z}{2(1-\rho^2)}\rbrack
\end{equation}
with
\begin{equation}
		z \equiv \frac{(x - \mu_{1})^2}{\sigma_{1}^2} - \frac{2\rho(x-\mu_{1})(y-\mu_{2})}{\sigma_{1}\sigma_{2}} + \frac{(y-\mu_{2})^2}{\sigma_2^2}
\end{equation}
Since we are working with a symmetric version of this formula where $\sigma_{1} = \sigma_{2}$ and $\mu_{1} = \mu_{2}$, we can rewrite this formula as
\begin{equation}
		P(x, y) = \frac{1}{2\pi\sigma^2\sqrt{1 - \rho^2}} exp \lbrack - \frac{z}{2(1-\rho^2)} \rbrack
\end{equation}
with
\begin{equation}
		z \equiv \frac{(x - \mu)^2 +(y - \mu)^2 - 2 \rho (x - \mu) (y - \mu)}{\sigma^2} \rbrack
\end{equation}
In Simpact, the following parameters are proposed: $\mu=4$ , $\sigma=1$ and $\rho=0.33$ as default values. 
This probability density function is limited on both axes to the interval [$\log_{10}4$ copies/ml, $\log_{10}8$ copies/ml].

The second model seeds new set-point viral loads from a log-Weibull distribution.
The set-point viral load in this case is calculated by taking the set-point viral load of the infector and adding Gaussian noise.

\begin{equation}
		\textbf{spvl} _{\mathit{infectee}}=\textbf{spvl} _{\mathit{infector}} + \mathcal{N}(\mu=0, \sigma^2=(\theta * \textbf{spvl} _{\mathit{infector}})^2)
\end{equation}

${\theta}$ determines the standard deviation of the noise to a fraction of the set-point viral load 
of the infector, it is 0.1 by default.\\
Since this addition of Gaussian noise can result in a negative set-point viral load. We resolve this by re-sampling the noise from the last step until the resulting set-point viral load is non-negative.

We propose a third way of modeling set-point viral load heritability, that uses the underlying set-point viral load distribution directly and results in a more intuitive and simpler model. The seeding of patients is not something that has a biological equivalent. Clearly, individuals do not develop HIV on their own without having had contact with another HIV-infected individual. Seeding events are introduced to bootstrap our simulations. We therefore believe that a plausible way of seeding an individual would be to sample a new set-point viral load value from a log-normal SPVL model. This way, seeding a new set-point viral load value is done by sampling one from the distribution
\begin{equation}
		\textbf{spvl}_{\mathit{infectee}}= LogNormal(\mu, \sigma^2)
\end{equation}

Transmission of a set-point viral load in this model is done by adding Gaussian noise to the set-point viral load of the infector:

\begin{equation}
		\textbf{spvl} _{\mathit{infectee}}=\textbf{spvl} _{\mathit{infector}} + \mathcal{N}(\mu=0, \sigma^2)
\end{equation}

\subsection{Exploring the SPVL model's parameter distributions}
For simulating local HIV transmission networks, like the ones found in Portugal, the set-point viral load transmission models need to be parameterized. More specifically, we are interested to learn the parameters' distribution. For this purpose, Markov chain Monte Carlo methods are often preferred. However, such methods require that a likelihood function is available. In the context of our set-point viral load models, such a likelihood function is computationally intractable. We therefore will use Approximate Bayesian Computation (ABC), as this technique allows the use of Bayesian inference, in a context where no likelihood function is available, but the output of a model can be compared to data using some kind of distance function.

In this work, we will compare the empirical distribution to the generated distribution using the Hellinger distance \cite{hellinger1909neue}.

\section{Experiments}
In this experiment, we use data that describes the Portuguese HIV-1 epidemic, and was exported from the RegaDB instance from the Egas Moniz hospital \cite{libin2013regadb}.

For our analyses, we require viral load samples and genetic sequences. The viral load samples are used to approximate the set-point viral load distribution. The genetic sequences are used to infer the transmission network. We use the polymerase genetic region of the HIV virus and more specifically consider the protease and reverse transcriptase regions. The reason for this is that these genetic regions are collected as standard patient care, as these regions are also used to assess drug resistance in HIV patients. In order to compute the distance matrix between the sequences, the sequences need to be aligned, for which we used the VIRULIGN method that implements a codon correct alignment procedure specifically tailored for virus sequences \cite{libin2018virulign}. As mentioned earlier, we observe that the genetic distance distribution significantly differs between subtypes, therefore the sequence dataset was split into three parts, subtype B (53\% of the dataset), subtype G (31\% of the dataset) and others. For this analysis, we will only consider the subtype B and subtype G datasets. The subtyping analysis was performed using the REGA genotyping framework \cite{Alcantara2009}. Note that in the context of transmission networks, separating between subtypes is a natural thing to do, as epidemics that are comprised of different subtypes are inherently separate.

From Figure \ref{fig:genetic_distance_per_subtype}, it is clear that the different subtypes require a custom threshold. For subtype B, we use the threshold as used in \cite{wertheim2013global}, and for subtype G we use the threshold that would yield the same percentile of the distribution as potential transmission partners.

We experimented with three different models of set-point viral load heritability, as defined in the previous section: a symmetric bi-variate model, a log-Weibull model with noise and a simple normal model. Furthermore, we explored two degree distributions: a global network, a network of Portuguese patients with subtype B (Waring distribution, $\rho = 3.1$) and a network of Portuguese patients with subtype G (Yule distribution, $\rho = 2.6$).
Additionally, we experimented with three different seeding proportions, corresponding to 0.27\%, 0.56\% and 0.83\% of all nodes.

\begin{figure}[H]
  \centering
  \includegraphics[width=0.4\linewidth]{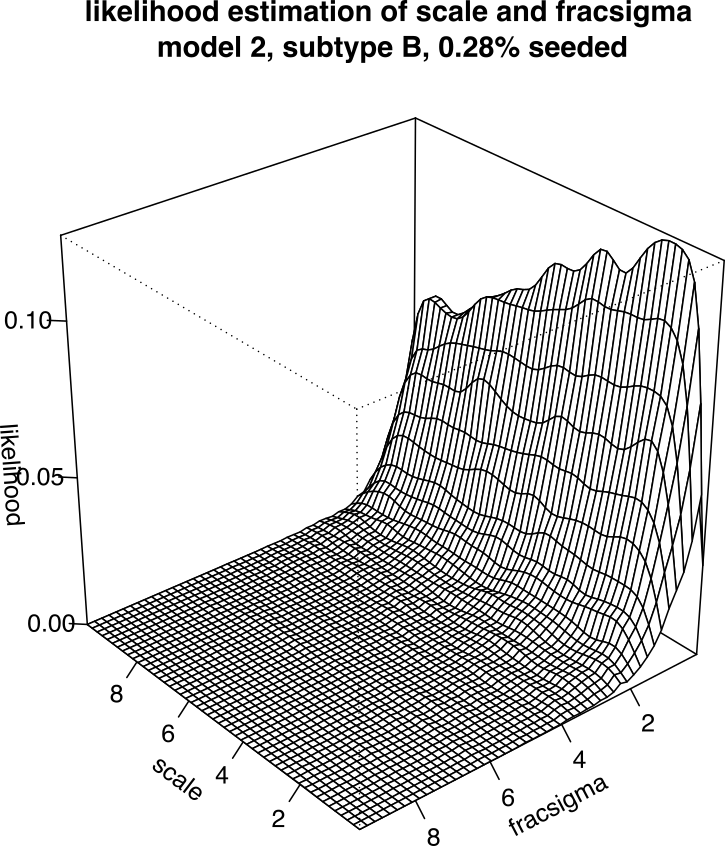}
  \includegraphics[width=0.4\linewidth]{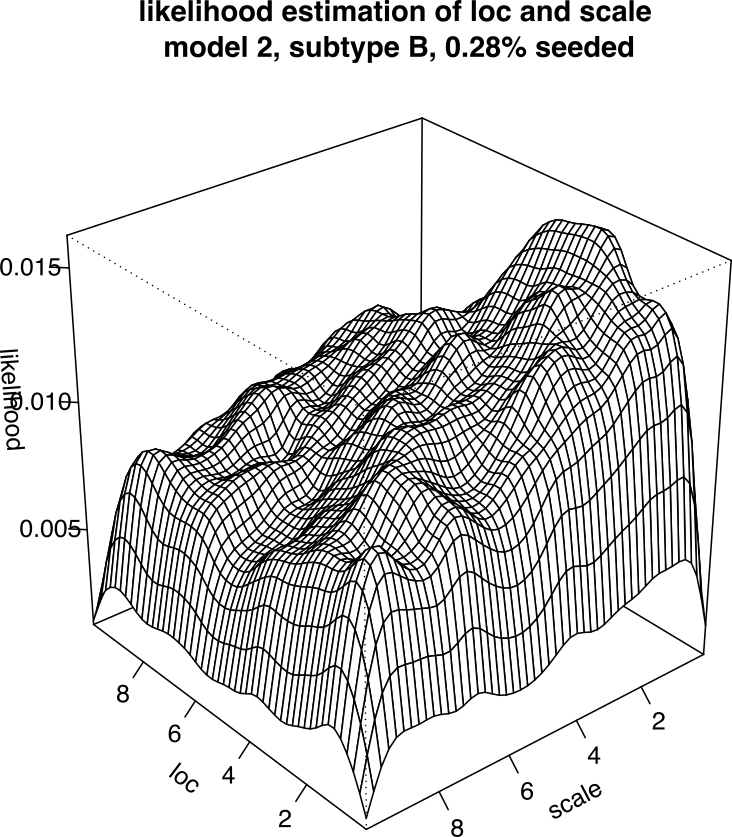}
  \includegraphics[width=0.4\linewidth]{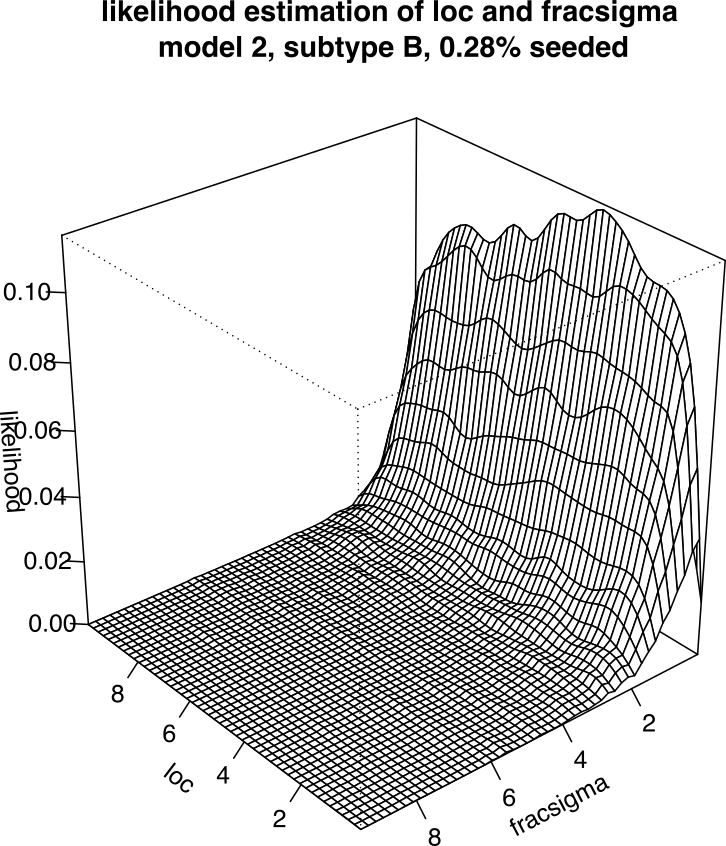}
  \caption{The posterior estimation of the parameters of the log-Weibull with noise model
    for the network with subtype B (sampled from a Waring distribution with $\rho = 3.1$) with 0.28\% of nodes seeded.
    In this figure, $loc$ and $scale$ are the parameters of the log-Weibull distribution and $fracsigma$ corresponds to the $\theta$ in subsection \ref{sec:models_of_hiv_spread}. 
\label{fig:posteriors}}
\end{figure}

\begin{figure}[H]
  \centering
  \includegraphics[width=0.4\linewidth]{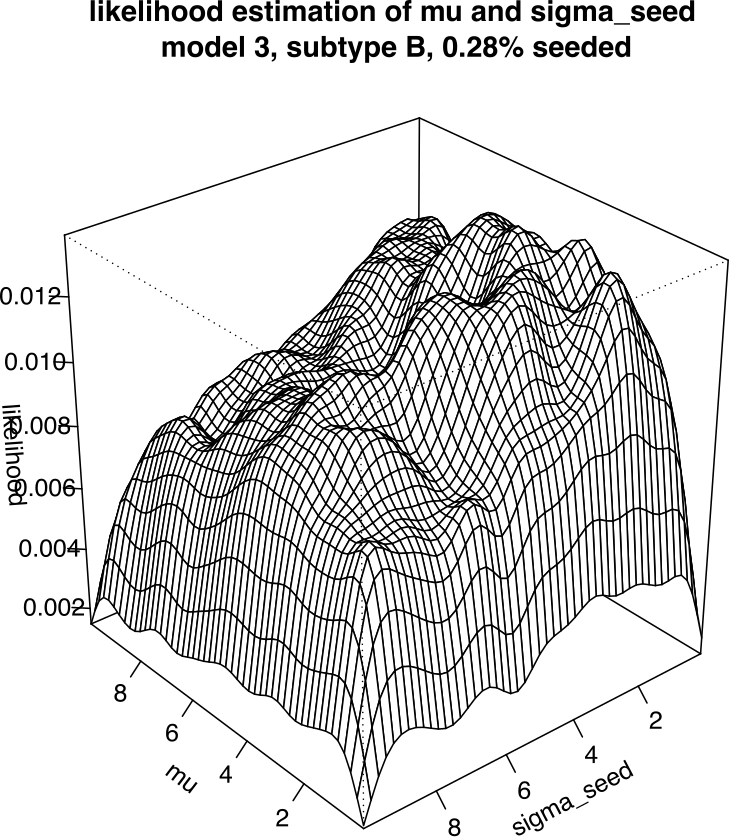}
  \includegraphics[width=0.4\linewidth]{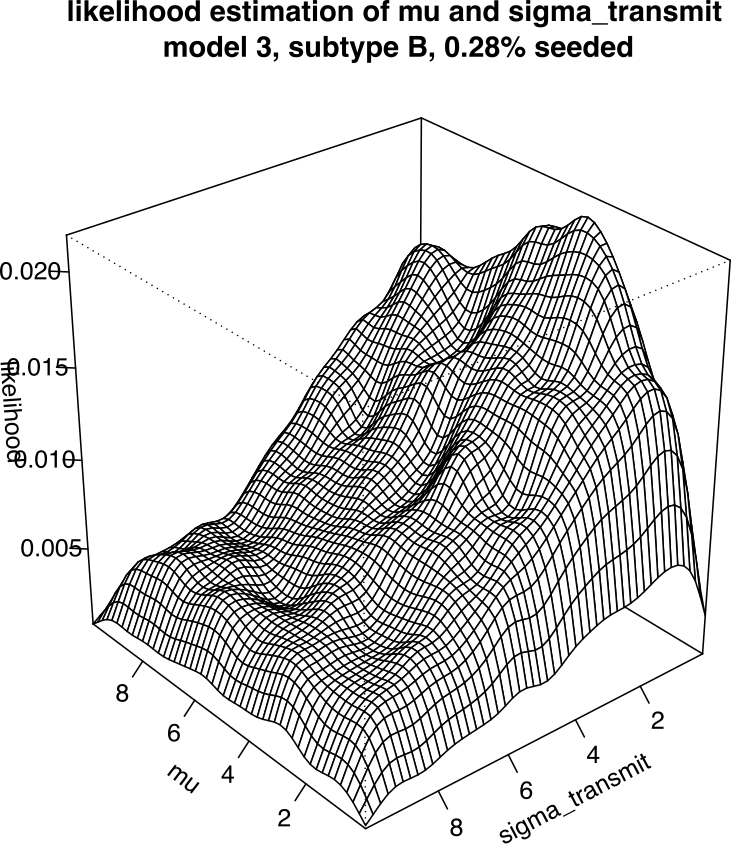}
  \includegraphics[width=0.4\linewidth]{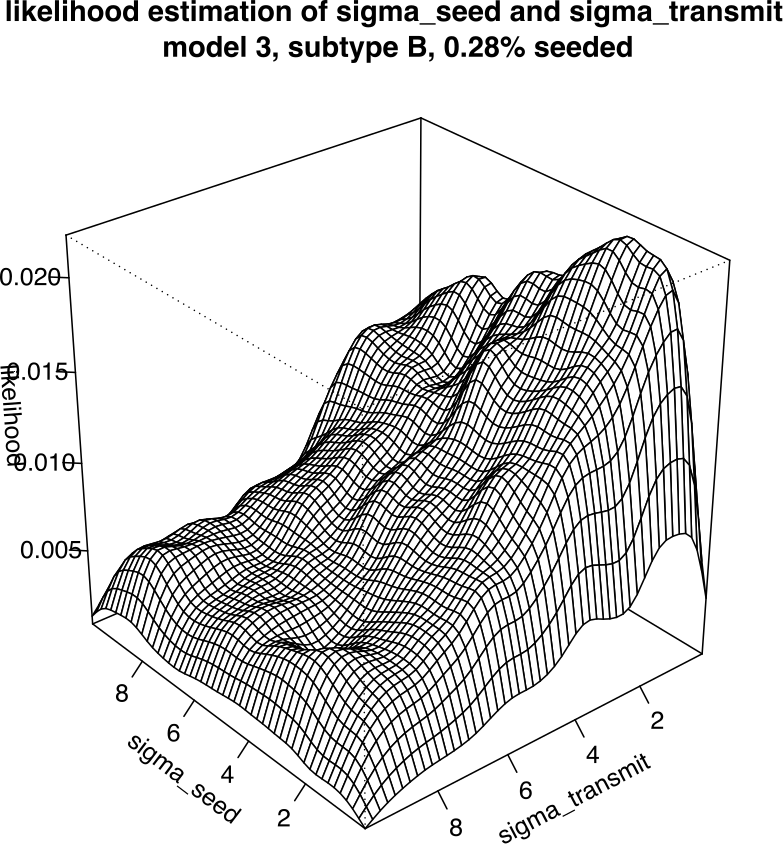}
  \caption{The posterior estimation of the parameters of the log-Normal model
    for the network with subtype B (sampled from a Waring distribution with $\rho = 3.1$) with 0.28\% of nodes seeded.
    In this figure, $mu$ and $sigma\_seed$ are the parameters of the log-Normal model and $sigma\_seed$ corresponds to the transmission noise parameter in subsection \ref{sec:models_of_hiv_spread}.
\label{fig:posteriors_ln}}
\end{figure}

We ran every ABC experiment for 200000 iterations and discarded the first 20000 ABC samples. An example result, from the log-Weibull model with noise for subtype B with 0.28\% seeded nodes, is shown in Figure \ref{fig:posteriors}, which demonstrates interesting parameter distributions (i.e., distinct peaks versus a more flat surface) that will be useful to inform prospective simulations in a more complex individual-based model.
Distinct peaks indicate clear models with little uncertainty, while flat surfaces indicate that the parameter might be less important, or that further analysis is warranted.
Another example result, from the log-Normal model for subtype B with 0.28\% seeded nodes, is shown in Figure \ref{fig:posteriors_ln}.

In future work, we plan to perform a detailed comparison between the different set-point viral load models we described, using the methods presented in this work, in conjunction with a more complex individual-based model.

\section{Conclusion and Discussion}
We present a new protocol that enables researchers to fit set-point viral load models in their local context, and diagnose the model parameter's uncertainty. Such parameter estimates are essential to enable subsequent modelling analyses, crucial to improve prevention policies.

For future work, we acknowledge that instead of using pairwise genetic distances, the inference of the transmission network could be improved by using viral distances inferred from a phylogenetic tree (i.e., patristic distances) \cite{libin_phylogeotool_2017}.

Currently we infer the parameters of the degree distribution as an external step using maximum likelihood estimation. In the future, we will investigate how the inference of the degree distribution can be integrated in the ABC process. 

\section*{Acknowledgments}
Pieter Libin was supported by a PhD grant of the FWO (Fonds Wetenschappelijk Onderzoek - Vlaanderen).
Kristof Theys was supported by a postdoctoral grant of the FWO.



\begin{thebibliography}{10}

\bibitem{who2017hivaids}
Hiv/aids, 2017.

\bibitem{cdc_hiv_report_2017}
Hiv/aids surveillance in europe, 2017.

\bibitem{national_port_2017}
Programa nacional para a infecao vih, sida e tuberculose, 2017.

\bibitem{Alcantara2009}
L.~C.~J. Alcantara, S.~Cassol, P.~Libin, K.~Deforche, O.~G. Pybus, M.~{Van
  Ranst}, B.~Galv{\~{a}}o-Castro, A.~M. Vandamme, and T.~de~Oliveira.
\newblock {A standardized framework for accurate, high-throughput genotyping of
  recombinant and non-recombinant viral sequences}.
\newblock {\em Nucleic Acids Research}, 37(SUPPL. 2), 2009.

\bibitem{alizon2010phylogenetic}
S.~Alizon, V.~von Wyl, T.~Stadler, R.~D. Kouyos, S.~Yerly, B.~Hirschel,
  J.~B{\"o}ni, C.~Shah, T.~Klimkait, H.~Furrer, et~al.
\newblock Phylogenetic approach reveals that virus genotype largely determines
  hiv set-point viral load.
\newblock {\em PLoS Pathog}, 6(9):e1001123, 2010.

\bibitem{chung2002connected}
F.~Chung and L.~Lu.
\newblock Connected components in random graphs with given expected degree
  sequences.
\newblock {\em Annals of combinatorics}, 6(2):125--145, 2002.

\bibitem{fraser2007variation}
C.~Fraser, T.~D. Hollingsworth, R.~Chapman, F.~de~Wolf, and W.~P. Hanage.
\newblock Variation in hiv-1 set-point viral load: epidemiological analysis and
  an evolutionary hypothesis.
\newblock {\em Proceedings of the National Academy of Sciences},
  104(44):17441--17446, 2007.

\bibitem{hellinger1909neue}
E.~Hellinger.
\newblock Neue begr{\"u}ndung der theorie quadratischer formen von
  unendlichvielen ver{\"a}nderlichen.
\newblock {\em Journal f{\"u}r die reine und angewandte Mathematik},
  136:210--271, 1909.

\bibitem{hodcroft2014contribution}
E.~Hodcroft, J.~D. Hadfield, E.~Fearnhill, A.~Phillips, D.~Dunn, S.~O'Shea,
  D.~Pillay, A.~J.~L. Brown, et~al.
\newblock The contribution of viral genotype to plasma viral set-point in hiv
  infection.
\newblock {\em PLoS Pathog}, 10(5):e1004112, 2014.

\bibitem{hool2013virus}
A.~Hool, G.~E. Leventhal, and S.~Bonhoeffer.
\newblock Virus-induced target cell activation reconciles set-point viral load
  heritability and within-host evolution.
\newblock {\em Epidemics}, 5(4):174--180, 2013.

\bibitem{lemey2009phylogenetic}
P.~Lemey.
\newblock {\em The phylogenetic handbook: a practical approach to phylogenetic
  analysis and hypothesis testing}.
\newblock Cambridge University Press, 2009.

\bibitem{Lemey2006}
P.~Lemey, A.~Rambaut, and O.~G. Pybus.
\newblock {HIV evolutionary dynamics within and among hosts}, 2006.

\bibitem{libin2013regadb}
P.~Libin, G.~Beheydt, K.~Deforche, S.~Imbrechts, F.~Ferreira, K.~Van~Laethem,
  K.~Theys, A.~P. Carvalho, J.~Cavaco-Silva, G.~Lapadula, et~al.
\newblock Regadb: community-driven data management and analysis for infectious
  diseases.
\newblock {\em Bioinformatics}, 29(11):1477--1480, 2013.

\bibitem{libin2018virulign}
P.~Libin, K.~Deforche, A.~B. Abecasis, and K.~Theys.
\newblock Virulign: fast codon-correct alignment and annotation of viral
  genomes.
\newblock {\em Bioinformatics}, page bty851, 2018.

\bibitem{libin_phylogeotool_2017}
P.~Libin, E.~{Vanden Eynden}, F.~Incardona, A.~Now{\'{e}}, A.~Bezenchek,
  A.~S{\"{o}}nnerborg, A.~M. Vandamme, K.~Theys, and G.~Baele.
\newblock {PhyloGeoTool: interactively exploring large phylogenies in an
  epidemiological context}.
\newblock {\em Bioinformatics (Oxford, England)}, 2017.

\bibitem{osmanov2002estimated}
S.~Osmanov, C.~Pattou, N.~Walker, B.~Schwardl{\"a}nder, and J.~Esparza.
\newblock Estimated global distribution and regional spread of hiv-1 genetic
  subtypes in the year 2000.
\newblock {\em Journal of acquired immune deficiency syndromes (1999)},
  29(2):184--190, 2002.

\bibitem{Palma2007}
A.~C. Palma, F.~Ara{\'{u}}jo, V.~Duque, F.~Borges, M.~T. Paix{\~{a}}o, and
  R.~Camacho.
\newblock {Molecular epidemiology and prevalence of drug resistance-associated
  mutations in newly diagnosed HIV-1 patients in Portugal}.
\newblock {\em Infection, Genetics and Evolution}, 2007.

\bibitem{pillay2007hiv}
D.~Pillay, A.~Rambaut, A.~M. Geretti, and A.~Brown.
\newblock Hiv phylogenetics.
\newblock {\em BMJ}, 335(7618):460--461, 2007.

\bibitem{tamura1993estimation}
K.~Tamura and M.~Nei.
\newblock Estimation of the number of nucleotide substitutions in the control
  region of mitochondrial dna in humans and chimpanzees.
\newblock {\em Molecular biology and evolution}, 10(3):512--526, 1993.

\bibitem{theys2010hiv}
K.~Theys.
\newblock Hiv-1 evolutionary dynamics: Individualized prediction of
  antiretroviral resistance.
\newblock Master's thesis, Katholieke Universiteit Leuven, 2010.

\bibitem{theys2018impact}
K.~Theys, P.~Libin, A.-C. Pineda-Pe{\~n}a, A.~Now{\'e}, A.-M. Vandamme, and
  A.~B. Abecasis.
\newblock The impact of hiv-1 within-host evolution on transmission dynamics.
\newblock {\em Current opinion in virology}, 28:92--101, 2018.

\bibitem{theys2009rise}
K.~Theys, J.~Vercauteren, A.~B. Abecasis, P.~Libin, K.~Deforche, A.-M.
  Vandamme, and R.~Camacho.
\newblock The rise and fall of k65r in a portuguese hiv-1 drug resistance
  database, despite continuously increasing use of tenofovir.
\newblock {\em Infection, Genetics and Evolution}, 9(4):683--688, 2009.

\bibitem{wang2016estimates}
H.~Wang, T.~M. Wolock, A.~Carter, G.~Nguyen, H.~H. Kyu, E.~Gakidou, S.~I. Hay,
  E.~J. Mills, A.~Trickey, W.~Msemburi, et~al.
\newblock Estimates of global, regional, and national incidence, prevalence,
  and mortality of hiv, 1980--2015: the global burden of disease study 2015.
\newblock {\em The lancet HIV}, 3(8):e361--e387, 2016.

\bibitem{wertheim2013global}
J.~O. Wertheim, A.~J.~L. Brown, N.~L. Hepler, S.~R. Mehta, D.~D. Richman, D.~M.
  Smith, and S.~L.~K. Pond.
\newblock The global transmission network of hiv-1.
\newblock {\em Journal of Infectious Diseases}, page jit524, 2013.

\end{thebibliography}
\end{document}